\documentclass[conference]{IEEEtran}

\usepackage{amsmath}
\usepackage{amssymb}

\usepackage{amsthm}
\newtheorem{theorem}{Theorem}

\usepackage{booktabs}
\usepackage{cite}
\usepackage[hidelinks]{hyperref}
\usepackage{multirow}
\usepackage{threeparttable}

\usepackage{algorithm}
\usepackage{algorithmic}

\ifCLASSOPTIONcompsoc
  \usepackage[caption=false,font=normalsize,labelfont=sf,textfont=sf]{subfig}
\else
  \usepackage[caption=false,font=footnotesize]{subfig}
\fi

\usepackage{tikz}
\usetikzlibrary{decorations.pathreplacing,fit,intersections}

\newcommand{\method}{ParaSeer}

\newcommand{\eqnref}[1]{\eqref{#1}}
\newcommand{\figref}[1]{\figurename~\ref{#1}}
\newcommand{\secref}[1]{Section~\ref{#1}}
\newcommand{\tableref}[1]{Table~\ref{#1}}
\newcommand{\thmref}[1]{Theorem~\ref{#1}}

\theoremstyle{definition}

\begin{document}

\title{Predicting Parameter Change's Effect \\ on Cellular Network Time Series}
\author{
    \IEEEauthorblockN{
        Mingjie Li\IEEEauthorrefmark{1}\IEEEauthorrefmark{4},
        Yongqian Sun\IEEEauthorrefmark{2},
        Xiaolei Hua\IEEEauthorrefmark{3},
        Renkai Yu\IEEEauthorrefmark{3},
        Xinwen Fan\IEEEauthorrefmark{3},
        Lin Zhu\IEEEauthorrefmark{3},
        Junlan Feng\IEEEauthorrefmark{3},
        Dan Pei\IEEEauthorrefmark{1}\IEEEauthorrefmark{4}
    }
    \IEEEauthorblockA{
        \IEEEauthorrefmark{1}Tsinghua University,
        \IEEEauthorrefmark{2}Nankai University,
        \IEEEauthorrefmark{3}China Mobile Research Institute,
    }
    \IEEEauthorblockA{
        \IEEEauthorrefmark{4}Beijing National Research Center for Information Science and Technology (BNRist)
    }
}

\maketitle

\setcounter{page}{1}
\thispagestyle{plain}
\pagestyle{plain}

\begin{abstract}
The cellular network provides convenient network access for ever-growing mobile phones.
During the continuous optimization, operators can adjust cell parameters to enhance the Quality of Service (QoS) flexibly.
A precise prediction of the parameter change's effect can help operators make proper parameter adjustments.
This work focuses on predicting cell status (like the workload and QoS) after adjusting the cell parameters.
The prediction will be conducted before an adjustment is actually applied to provide an early inspection.
As it can be hard for available parameter adjustments with a limited number to cover all the parameter and user behavior combinations, we propose \textit{\method{}} fusing domain knowledge on parameter adjustments into data-driven time series forecasting.
\method{} organizes several pre-trained Transformers for adjustment-free time series forecasting, utilizing plenty of adjustment-free data.
On the other hand, \method{} models the effect of adjusting the transmission power and cell individual offset (CIO) as a multiplier for the workload.
We derive a formula to calculate the multiplier from the underlying mechanism of those two parameters, helping \method{} eliminate the thirst for data with parameter adjustments.
We compare \method{} with baselines on two real-world datasets, where \method{} outperforms the best baseline by more than 25.8\% in terms of RMSE.
The extensive experiments further illustrate the contributions of \method{}'s components.
\end{abstract}



\section{Introduction}\label{sec:introduction}

The cellular network enables network access for user equipment (UE), \textit{e.g.}, mobile phones, smartwatches, and portable hotspots.
With 5G in view, this service is being established on a large number of next generation NodeBs (gNBs), which are carefully located by cellular service providers to optimize their workload and Quality of Service (QoS).
Through continuous optimization, cellular networks can provide satisfactory service almost all the time.
Nevertheless, there may still be times when the service is unsatisfactory, such as unexpected high usage or a scheduled shutdown of some gNBs for maintenance.
Fortunately, operators can optimize network conditions by adjusting configuration and optimization parameters like \textit{transmission power} and \textit{cell individual offset} (\textit{CIO})~\cite{Shodamola:2020,Alsuhli:2021}, which are the two most frequently adjusted parameters in our scenario.
Currently, expert operators configure parameters manually based on their experience~\cite{Klaine:2017}, which is not always effective.
Predicting the parameter change's effect in advance can help operators make proper parameter adjustments.

In this work, we focus on predicting cell status (like the workload and QoS) after a given cell parameter adjustment, providing an early inspection before the real application for operators.
Cells in the cellular network present complex spatial-temporal dependencies as UEs migrate among gNBs~\cite{Wang:2021}.
It is impractical to model the whole history of all the cells at the same time.
As a result, it is common to slice time series into windows~\cite{Xu:2018,Wu:2021,Jiang:2022}.
As for spatial dependency, we focus on a single cell without data from its neighbors for simplicity.
We leave modeling multiple cells and recommending parameters to achieve the desired status as future work.

\begin{figure}[t]
	\centering
    \includegraphics[width=0.95\columnwidth]{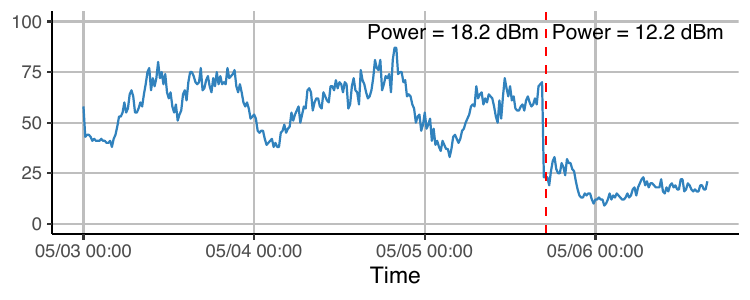}
	\caption{
        Time series of the maximum number of RRC connections around an adjustment from a real-world cell.
        The dashed vertical line indicates the first data point with the new transmission power.
    }\label{fig:example}
\end{figure}

\figref{fig:example} shows the time series of a cell's maximum number of Radio Resource Control (RRC) connections from a real-world cellular network.
Before lowering the transmission power (on the left of the vertical line in \figref{fig:example}), the time series presents the periodicity of the cell's users.
As operators lowered the transmission power by 6 dBm (decibel-milliwatt), cell edge UEs were handed over to the neighboring cells for higher reference signal received power (RSRP), which led to a dramatic drop in the time series.
The time series on the right of the vertical line in \figref{fig:example}, as well as other monitoring metrics (\textit{i.e.}, multivariate sequences), is the interest of this work.

Traffic prediction~\cite{Jiang:2022,Lin:2022} and general time series forecasting~\cite{Zeng:2023} are well-studied in the literature.
To the best of our knowledge, none of the existing traffic prediction works in the network community considers the adjustments of cell parameters.
As shown in \figref{fig:example}, the data pattern varies after the parameter adjustment, \textit{i.e.}, a \textit{concept drift}~\cite{Gama:2014} arose.
Notice that concept drift adaptation relies on observation of the new concept~\cite{Ma:2018}, which is unavailable in our task.
We further relate our problem with \textit{causal effect estimation} from observational data~\cite{Zhang:2021:TESurvey}, \textit{i.e.}, how adjusting parameters will influence another variable, like workload.
As the difference among diverse parameter adjustments is not the focus of this work, our task reduces to a regression problem from the historical multivariate sequences and an adjustment to the following data points (the single-model approaches in \cite{Zhang:2021:TESurvey}).

Predicting parameter change's effect confronts two main challenges.
The first challenge comes from \textit{the lack of more data with cell parameter adjustments}.
Intuitively, with sufficient historical cases available, we can utilize big data to mine the patterns of parameter changes' effect, \textit{e.g.}, learning how the transmission power adjustment will influence the maximum number of RRC connections from \figref{fig:example}.
However, there are not many historical examples of parameter adjustments, and they cannot cover all scenarios.
For example, the transmission power in the collaborating cellular network can be up to 49 dBm.
There will be hundreds of options with a resolution of 0.1 dBm.
Let alone 2000/site parameters for 5G~\cite{Shodamola:2020}.

Instead of directly predicting cell status after parameter adjustments, we plan to implement it in two phases to address the first challenge.
In the first phase, we assume that the parameters remain unchanged and predict the trend of the cell status.
In the second phase, we predict how the cell status will shift when the parameters are adjusted.

There have been many studies on time series forecasting for the first phase.
However, the second phase confronts the challenge of \textit{modeling precisely the underlying data generation process} because a data-driven regression model may suffer from performance degradation with limited data due to its incompatibility with the underlying data generation process.
For example, a linear model cannot handle nonlinear relations well.

In this work, we propose \textit{\method{}}, addressing the above challenges.
We first pre-train a time series forecasting model, utilizing plenty of adjustment-free data.
Considering the complex relationships among metrics, we adopt Transformer~\cite{Vaswani:2017} as its backbone, which is believed to be a universal approximator of sequence-to-sequence functions~\cite{Yun:2020}.
After that, \method{} modifies the adjustment-free prediction to approach the change's effect.
To address the second challenge, we introduce the graphical model~\cite{Bareinboim:2022} to qualitatively model the propagation of parameter change's effect among monitoring variables.
Specifically speaking, \method{} organizes multiple Transformers for each monitoring variable in the graphical model and embeds causal parents into a Transformer's input to propagate the parameter change's effect along the graphical model.
Inspired by the fact that both the transmission power and CIO determine the boundary of a cell, we further quantitatively model the effect of adjusting these two parameters by modeling the change in the cell area as a multiplier.
A formula is derived for calculating the multiplier, which will be applied to adjustment-free prediction for the workload.
The encoded domain knowledge helps \method{} eliminate the thirst for data with parameter adjustments.

In a real-world dataset from our cellular service provider partner, the proposed \method{} outperforms the best baseline by 25.8\% and 59.0\% in terms of RMSE and sMAPE.
The ablation studies further illustrate the positive effect of \method{}'s components.
Moreover, we find that further fine-tuning is unnecessary for \method{}, and \method{}'s performance degradation for new cells is insignificant.
Another dataset from a different geographical region verifies \method{}'s advantage over baselines, implying \method{}'s generalization potential for new cells.
We summarize our contributions as follows.

\begin{itemize}
    \item We propose \method{} to predict the parameter change's effect on cellular network time series, intending to help operators make proper parameter adjustments.
    \method{} decomposes prediction into two sub-tasks, \textit{i.e.}, 1) adjustment-free time series forecasting and 2) modifying the adjustment-free prediction.
    For the first sub-task, \method{} adopts the prevalent Transformer as its backbone.
    On the other hand, \method{} models the parameter change's effect through the change in the cell area as a multiplier for cell workload.
    \item We derive a formula to calculate the multiplier from the underlying mechanism of the transmission power and CIO.
    The encoded domain knowledge helps \method{} eliminate the thirst for data with cell parameter adjustments.
    Moreover, the formula's universal derivation guarantees \method{}'s explainability.
    \item We extensively evaluate \method{} on two real-world datasets from different geographical regions.
    Experiment results show that \method{} outperforms baselines by more than 25.8\% and 59.0\% in terms of RMSE and sMAPE, respectively.
    The ablation study further illustrates the effect of \method{}'s components.
\end{itemize}

The rest of this paper is organized as follows.
\secref{sec:background} provides more background on the cellular network, followed by the problem statement.
We propose a homogeneous single-cell model in \secref{sec:hsc}, based on which \secref{sec:methodology} presents the detail of \method{}.
We evaluate \method{} and baselines with two real-world datasets from our cellular service provider partner in \secref{sec:experiment}.
We discuss related works in \secref{sec:relate-works} and conclude in \secref{sec:conclusion}.


\section{Background}\label{sec:background}

\subsection{Parameters in the Cellular Network}\label{sec:background:parameters}

A gNB usually has more than one antenna, each serving a region named a \textit{cell}.
Operators can configure cells separately.
Besides some common parameters from the gNB, \textit{e.g.}, the latitude and longitude, each cell has its sole parameters, \textit{e.g.}, the mechanical tilt of the corresponding antenna.
The transmission power is a fundamental parameter of a cell.
The higher the transmission power, the higher the RSRP.
A UE prefers the cell with the highest RSRP.
3GPP introduces the CIO parameter to control handover for proper load balancing~\cite{3gpp.38.331}.
Following the 3GPP standard specification, a UE selects the target cell $t$ if it provides better RSRP than the source cell $s$.
Equation~\eqnref{eqn:cell-selection} formulates the criterion as an example, where $Hys$ is the hysteresis parameter to reduce handover for cell edge UEs.
As a result, CIO and $Hys$ influence when a UE is handed over from one cell to a neighboring one.

\begin{equation}\label{eqn:cell-selection}
    RSRP_{t} + CIO_{t \to s} > RSRP_{s} + CIO_{s \to t} + Hys
\end{equation}

This work focuses on the transmission power and CIO.
Both parameters play an important role in the aforementioned cell selection process, determining the boundary of a cell, \textit{i.e.}, where UEs can obtain service from a given cell.

\subsection{Monitoring}

\begin{table*}[t]
    \centering
    \caption{Clusters of monitoring metrics}\label{tab:variables}
    \begin{tabular}{l c p{0.25\linewidth} p{0.53\linewidth}}
        \toprule
        Cluster & Symbol & Description & Monitoring Metrics \\
        \midrule
        Workload & $\mathbf{W}$ & The number and behaviors of served UEs. & \#(RRC connection established), \#(E-RAB connection established), PRB downlink / uplink utilization, maximum \#(RRC connection), average \#(RRC connection), CCE utilization in PDCCH \\
        Interference & $\mathbf{I}$ & The strength of irrelevant electromagnetic waves. & average noise level of the uplink PRBs \\
        QoS & $\mathbf{Q}$ & Quality of Service measurements. & drop rate, connection success rate, RRC connection success rate, E-RAB connection success rate, E-RAB drop rate (QCI=1), E-RAB connection success rate (QCI=1), handover success rate, VOLTE handover success rate, paging congestion rate \\
        \bottomrule
    \end{tabular}
\end{table*}

A gNB gathers the running status for each cell, which is further summarized as time series with a constant frequency, \textit{e.g.}, every 15 minutes.
The monitoring metrics collected from cells of a nationwide cellular network can be clustered into several groups, as shown in \tableref{tab:variables}.
Some metrics measure the workload, \textit{e.g.}, how many connections have been established.
Another kind of metrics describes a cell's QoS, \textit{e.g.}, drop rate and connection success rate.
Moreover, some factors are out of a gNB's control and can also worsen QoS, measured as the noise level.

There is a group of handover-related monitoring metrics, counting the number of UEs handed over to a neighboring cell.
Once a parameter is adjusted (\textit{e.g.}, lowering the transmission power), the handover counters change dramatically in the first data point and back to normal in the following ones.
In other words, such process quantities behave differently from state quantities like workload and QoS.
Hence, we filter handover-related monitoring metrics out of consideration as future work.

\subsection{Graphical Model}

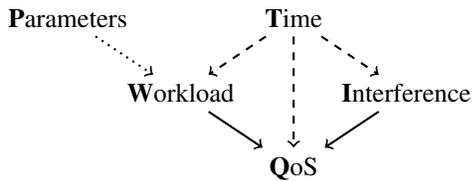
\begin{figure}[t]
    \centering

\begin{tikzpicture}[
	node distance=3cm,
	align=center,
	execute at begin node=\setlength{\baselineskip}{10pt}
	]
	\node(workload){\textbf{W}orkload};
	\node[right of=workload](interference){\textbf{I}nterference};
	\node[right of=workload, xshift=-1.5cm, yshift=-1cm](qos){\textbf{Q}oS};
	\node[right of=workload, xshift=-1.5cm, yshift=1cm](time){\textbf{T}ime};
	\node[left of=workload, xshift=1.5cm, yshift=1cm](parameters){\textbf{P}arameters};

    \draw[thick, dotted, ->] (parameters) -- (workload);
    \draw[thick, ->] (workload) -- (qos);
    \draw[thick, ->] (interference) -- (qos);
    \draw[thick, dashed, ->] (time) -- (workload);
    \draw[thick, dashed, ->] (time) -- (interference);
    \draw[thick, dashed, ->] (time) -- (qos);
\end{tikzpicture}
    \caption{
        Graphical model for a single cell.
        Dashed arrows from time encode the periodicity of the related variables.
    }\label{fig:graphical-model}
\end{figure}

We present the relations among clustered monitoring metrics with a graphical model~\cite{Bareinboim:2022}, as shown in \figref{fig:graphical-model}.
Each node in the graph is a cluster of monitoring metrics or parameters, named \textit{variables}.
Links among variables encode our understanding of the underlying data generation mechanism.

We introduce a variable named \textit{time} with dashed arrows to represent the periodicity of monitoring metrics.
Specifically, we extract the day of the week, the hour of the day, and the minute from timestamps as time series.
Because the direct effect of the transmission power and CIO is the boundary of a cell, we draw an arrow from \textit{parameters} to \textit{workload} in the graphical model.
Meanwhile, the missing link between \textit{parameters} and \textit{QoS} indicates an indirect relation.

In the rest of this paper, we use a bold uppercase letter to represent a variable, \textit{e.g.}, $\mathbf{W}$ represents \textbf{W}orkload.
Boldface means $\mathbf{W}$ may be a set of elements, \textit{i.e.}, a cluster of monitoring metrics.
A lowercase letter (\textit{e.g.}, $\mathbf{w}$) represents one of the potential values of the corresponding variable.
Moreover, we denote the parents of a variable $\mathbf{X}$ in the graphical model as $\mathbf{Pa}(\mathbf{X})$, \textit{e.g.}, $\mathbf{Pa}(\mathbf{Q}) = \{\mathbf{W}, \mathbf{T}, \mathbf{I}\}$.
The timestamp will be labeled as a superscript for elements in a time series.

\subsection{Problem Statement}

Denote multivariate time series as $\left(\mathbf{x}^{(t - I)}, \cdots, \mathbf{x}^{(t - 1)}\right)$, where $\mathbf{x}^{(t - I)} \in \mathbb{R}^{d_{\mathbf{X}}}$ is a vector of monitoring metrics sampled at time $t - I$ and $I$ is the look-up window as \textbf{i}nput.
Given that the parameters are adjusted by $\Delta\mathbf{p} \in \mathbb{R}^{d_{\mathbf{P}}}$ at time $t$, we aim to predict the following monitoring metrics $\left(\mathbf{x}^{(t)}, \cdots, \mathbf{x}^{(t + O - 1)}\right)$.

This work focuses on single-cell models, as mentioned in \secref{sec:introduction}.
We leave a more precise multi-cell model as future work.
Moreover, a natural follow-up work is to recommend $\Delta\mathbf{p}$ for desired cell status.

\section{Homogeneous Single-Cell Modeling}\label{sec:hsc}

\subsection{Motivation: Modeling Workload by Area}

Cell edge UEs are sensitive to an adjustment of the transmission power or CIO.
Given a cell edge UE $U$ and its corresponding cell $A$, after decreasing $A$'s transmission power (or CIO), $U$'s RSRP may satisfy \eqnref{eqn:cell-selection} and trigger a handover to a neighboring cell $B$, disconnecting $U$ from $A$.
As a result, $A$'s workload decreases by $U$'s usage, while $B$'s workload increases.

The change in the workload may be different if $U$'s owner is using another UE at the same time.
Her two UEs may share the same RSRP as they are in the same position.
Hence, the same adjustment may lead to a double decrease in $A$'s workload.
As a result, we introduce the concept of \textit{usage density}, modeling the workload by integrating the usages over the area, \textit{i.e.}, $\mathbf{W} = \iint_{\mathbf{S}} \mathbf{\rho}(x,y) \,\mathrm{d}x\,\mathrm{d}y$, where $\mathbf{S}$ is $A$'s serving region and $\mathbf{\rho}(x,y)$ is the usage density function.

We argue that we can infer all the UE trajectories with an ideal solution as implied by \thmref{thm:track-necessity} (proved in the Appendix), where the gNBs' serving regions are represented by the multiplicatively weighted Voronoi diagram~\cite{Portela:2006}.
In other words, \textit{it is mandatory for a perfect prediction to take fine-grained UE trajectories as input (at least during training to some extent), which we will never collect due to privacy considerations}.
Hence, we decide to develop a trajectory-free model at the cost of accuracy.
As a result, our best estimation is that the cell edge shares the same average usage density as the whole cell, named the \textit{homogeneous usage density assumption}.

\begin{theorem}\label{thm:track-necessity}
Let $\mathbf{c}_{i} \in \mathbb{R}^{2}$ be the $i$-th site point in the plane, where $i = 1, 2, \cdots$.
Denote the Voronoi region generated by $\mathbf{c}_{i}$ as
$\mathbf{S}_{i} = \{ \mathbf{x} \mid \forall (j \neq i) \frac{\lVert \mathbf{x} - \mathbf{c}_{i} \rVert}{\phi_{i}} \le \frac{\lVert \mathbf{x} - \mathbf{c}_{j} \rVert}{\phi_{j}}, \mathbf{x} \in \mathbb{R}^{2} \}$,
where $\phi_{i} > 0$ and $\lVert \mathbf{x} - \mathbf{c}_{i} \rVert$ is the distance between $\mathbf{x}$ and $\mathbf{c}_{i}$.
For any given $\Phi = \{ \phi_{i} \mid i = 1, 2, \cdots\}$, a region's area is finite, \textit{i.e.}, $\lVert \mathbf{S}_{i} \rVert < +\infty$.

Given a density function $\rho$ defined over the plane $\mathbb{R}^{2}$ and a region-specific accumulator $f_{i}(\Phi) = \iint_{\mathbf{S}_{i}} \rho(x,y) \,\mathrm{d}x\,\mathrm{d}y$, we can reconstruct $\rho$ from $F = \{f_{i} \mid i = 1, 2, \cdots\}$.
\end{theorem}

Under the homogeneous usage density assumption, our core idea is to estimate the change in the cell area $\frac{\lVert \mathbf{S}^{\prime} \rVert}{\lVert \mathbf{S} \rVert}$ as a proxy for the change in the workload.
After that, we can conduct adjustment-free time series forecasting for the workload and further infer the other monitoring metrics under the adjusted workload.

\subsection{Cell Boundary Determination}

The hysteresis parameter in \eqnref{eqn:cell-selection} extends the cell boundary into a stripe with a width.
It will be convenient to find a virtual boundary within which the number of UEs connected to other cells is the same as that on the other side but connected to the concerned cell.
With the area bounded by such a virtual boundary, we can omit the hysteresis parameter's effect on the UE number and other workload measurements.

Intuitively, we calculate a virtual boundary between two cells where a UE observes the same offset RSRPs, \textit{i.e.}, solving \eqnref{eqn:virtual-boundary}.
Notice that the parameters (\textit{e.g.}, $CIO_{t \to s}$) of neighboring cells also influence the cell boundary.
As a result, our single-cell model rests on the \textit{homogeneous parameter assumption}, \textit{i.e.}, neighboring cells share the same parameters as those of the cell under discussion.

\begin{equation}\label{eqn:virtual-boundary}
    RSRP_{t} + CIO_{t \to s} = RSRP_{s} + CIO_{s \to t}
\end{equation}

The transmission power and CIO takes ``dBm'' as the unit in gNB operations.
When the transmission power changes from $\tau$ dBm to $\tau + \delta$ dBm and CIO changes from $o$ dB to $o + \delta_{o}$ dB, a UE connected to the cell will observe a change in the offset RSRP from $P_{R}$ to $P_{R}^{\prime} = \beta P_{R}$, where $\beta = 10^{0.1 (\delta + \delta_{o})}$.
Under the homogeneous parameter assumption, $\beta$ is also the ratio between the concerned cell's offset transmission power and that of its neighbors.

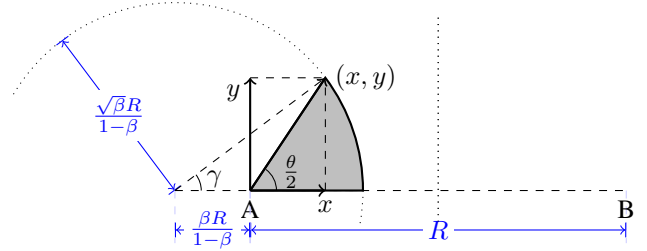
\begin{figure}[t]
	\centering

\begin{tikzpicture}[
	node distance=3cm,
	align=center,
	execute at begin node=\setlength{\baselineskip}{10pt}
	]
	\draw[draw, thick, fill=gray!50] (0, 0) -- (1.5cm, 0) arc[start angle=0, end angle=36.9, radius=2.5cm] -- (0, 0);

	\draw[dotted, black] (2.5cm, 2.3cm) -- (2.5cm, -0.45cm);
    \begin{scope}[blue, inner sep=2pt]
        \draw[dashed, blue!20] (0, 0) -- +(0, -0.7cm)
            node [coordinate, near end] (a) {};
        \draw[dashed, blue!20] (5cm,0) -- +(0, -0.7cm)
            node [coordinate, near end] (b) {};
        \draw[|<->|] (a) -- node[fill=white] {$R$} (b);
    \end{scope}
	\draw[dashed] (0cm, 0cm) node[below]{A} -- (5cm, 0) node[below]{B};

	\draw[dotted, black] (1.5cm, 0) arc[start angle=0, end angle=150, radius=2.5cm];
	\draw[dotted, black] (1.5cm, 0) arc[start angle=0, end angle=-10, radius=2.5cm];
	\draw[|<->|, blue, inner sep=2pt] (-1cm, 0) -- node[fill=white]{$ \frac{\sqrt{\beta} R}{1-\beta}$} +(-1.5cm, 2cm);
	\draw[dotted, black] (2.5cm, 2.3cm) -- (2.5cm, -0.45cm);
    \begin{scope}[blue, inner sep=2pt]
        \draw[dashed, blue!20] (0, 0) -- +(0, -0.7cm)
            node [coordinate, near end] (a) {};
        \draw[dashed, blue!20] (-1cm,0) -- +(0, -0.7cm)
            node [coordinate, near end] (b) {};
        \draw[|<->|] (a) -- node[fill=white] {$\frac{\beta R}{1 - \beta}$} (b);
    \end{scope}

	\draw[->, draw, thick] (0, 0) node[right, yshift=0.25cm, xshift=0.3cm]{$\frac{\theta}{2}$} -- (1cm, 1.5cm) node[right]{$(x, y)$};
	\draw[->, draw, thick] (0, 0) -- (1cm, 0) node[below]{$x$};
	\draw[->, draw, thick] (0, 0) -- (0, 1.5cm) node[left, yshift=-0.2cm]{$y$};
	\draw[dashed, draw] (1cm, 0) -- (1cm, 1.5cm);
	\draw[dashed, draw] (0, 1.5cm) -- (1cm, 1.5cm);
	\draw (0.35cm, 0) arc[start angle=0, end angle=53.1, radius=0.35cm, thick];

	\draw[draw, dashed] (-1cm, 0) node[right, yshift=0.15cm, xshift=0.3cm]{$\gamma$} -- (1cm, 1.5cm);
	\draw[draw, dashed] (-1cm, 0) -- (1.5cm, 0);
	\draw (-0.65cm, 0) arc[start angle=0, end angle=36.9, radius=0.35cm, thick];
\end{tikzpicture}
	\caption{
        Two omnidirectional cells.
        As $A$'s transmission power changes from $\tau$ dBm to $\tau + \delta$ dBm and its CIO changes from $o$ dB to $o + \delta_{o}$ dB, a UE connected to $A$ will observe a change in the offset RSRP from $P_{R}$ to $P_{R}^{\prime} = \beta P_{R}$, where $\beta = 10^{0.1 (\delta + \delta_{o})}$.
        With $\beta < 1$, $A$'s boundary shrinks to the dotted circle.
    }\label{fig:bi-antenna}
\end{figure}

Notice that the power density of electromagnetic waves decays quadratically with the distance from the point source.
Hence, the coordinates $(x, y)$ of a UE on the virtual border between two neighboring omnidirectional cells satisfy \eqnref{eqn:bi-antenna-boundary} according to \eqnref{eqn:virtual-boundary}, where $P$ is the initial transmission power, and $R$ is the distance between these two gNBs.
When $\beta \neq 1$, \eqnref{eqn:bi-antenna-boundary} further provides \eqnref{eqn:bi-antenna-boundary:circle}, known as the \textit{Apollonius circle}.
\figref{fig:bi-antenna} visualizes the related quantities for $\beta < 1$.
As for sectored cells, we treat the border between two neighboring cells of the same gNB as a ray starting from the gNB.

\begin{equation}\label{eqn:bi-antenna-boundary}
    \underbrace{\frac{P}{(R - x)^{2} + y^{2}}}_{RSRP_{t}}
    \cdot \underbrace{\vphantom{\frac{10^{\delta}}{y^{2}}} 10^{0.1 o}}_{CIO_{t \to s}}
    = \underbrace{\frac{10^{0.1 \delta} \cdot P}{x^{2}+y^{2}}}_{RSRP_{s}}
    \cdot \underbrace{\vphantom{\frac{10^{\delta}}{y^{2}}} 10^{0.1 (o + \delta_{o})}}_{CIO_{s \to t}}
\end{equation}

\begin{equation}\label{eqn:bi-antenna-boundary:circle}
    \left(x + \frac{\beta R}{1 - \beta} \right)^{2} + y^{2} = \frac{\beta R^{2}}{(1-\beta)^{2}}, \beta \neq 1 \wedge \beta > 0
\end{equation}

\begin{figure}[t]
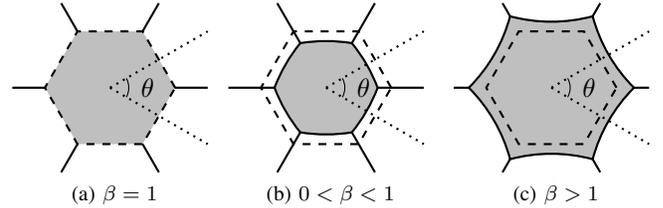

    \centering
	\subfloat[$\beta=1$]{

\begin{tikzpicture}[
	align=center,
	execute at begin node=\setlength{\baselineskip}{10pt}
	]
    \input{img/cells-constant}

    \draw[thick] (site-0) -- (vertex-1-1);
    \draw[thick] (site-0) -- (vertex-2-2);
    \draw[thick] (site-0) -- (vertex-3-3);
    \draw[thick] (site-0) -- (vertex-4-4);
    \draw[thick] (site-0) -- (vertex-5-5);
    \draw[thick] (site-0) -- (vertex-6-6);
    \draw[draw, thick, dashed, fill=gray!50] (vertex-1) -- (vertex-2) -- (vertex-3) -- (vertex-4) -- (vertex-5) -- (vertex-6) -- (vertex-1);

    \draw[draw, dotted, thick] (site-0) -- (site-1);
    \draw[draw, dotted, thick] (site-0) -- (site-6);
	\node at (0.5cm, 0) {$\theta$};
	\draw (0.25cm * \SqrtThree / 2, 0.25cm / 2) arc[start angle=\THETA / 2, end angle=-\THETA / 2, radius=0.25cm, thick];
\end{tikzpicture}}
	\subfloat[$0 < \beta < 1$]{

\begin{tikzpicture}[
	align=center,
	execute at begin node=\setlength{\baselineskip}{10pt}
	]
    \input{img/cells-constant}

    \draw[thick] (site-0) -- (vertex-1-1);
    \draw[thick] (site-0) -- (vertex-2-2);
    \draw[thick] (site-0) -- (vertex-3-3);
    \draw[thick] (site-0) -- (vertex-4-4);
    \draw[thick] (site-0) -- (vertex-5-5);
    \draw[thick] (site-0) -- (vertex-6-6);

    \draw[draw, thick, fill=gray!50] (\CBSD cm * \SqrtThree / 2, \CBSD cm * 0.5)
        arc[start angle=30, end angle=30+\GAMMA, radius=\RADIUS cm]
        arc[start angle=90-\GAMMA, end angle=90+\GAMMA, radius=\RADIUS cm]
        arc[start angle=150-\GAMMA, end angle=150+\GAMMA, radius=\RADIUS cm]
        arc[start angle=210-\GAMMA, end angle=210+\GAMMA, radius=\RADIUS cm]
        arc[start angle=270-\GAMMA, end angle=270+\GAMMA, radius=\RADIUS cm]
        arc[start angle=-30-\GAMMA, end angle=-30+\GAMMA, radius=\RADIUS cm]
        arc[start angle=30-\GAMMA, end angle=30, radius=\RADIUS cm];

    \draw[draw, thick, dashed] (vertex-1) -- (vertex-2) -- (vertex-3) -- (vertex-4) -- (vertex-5) -- (vertex-6) -- (vertex-1);

    \draw[draw, dotted, thick] (site-0) -- (site-1);
    \draw[draw, dotted, thick] (site-0) -- (site-6);
	\node at (0.5cm, 0) {$\theta$};
	\draw (0.25cm * \SqrtThree / 2, 0.25cm / 2) arc[start angle=\THETA / 2, end angle=-\THETA / 2, radius=0.25cm, thick];
\end{tikzpicture}
        \label{fig:multi-cells:lt}
    }
	\subfloat[$\beta > 1$]{

\begin{tikzpicture}[
	align=center,
	execute at begin node=\setlength{\baselineskip}{10pt}
	]
    \input{img/cells-constant}

    \draw[thick] (site-0) -- (vertex-1-1);
    \draw[thick] (site-0) -- (vertex-2-2);
    \draw[thick] (site-0) -- (vertex-3-3);
    \draw[thick] (site-0) -- (vertex-4-4);
    \draw[thick] (site-0) -- (vertex-5-5);
    \draw[thick] (site-0) -- (vertex-6-6);

    \pgfmathsetmacro{\BIGBETA}{1 / \BETA}
    \pgfmathsetmacro{\INTERSECTION}{(\BIGBETA * cos(\THETA / 2) - sqrt(\BIGBETA * (1 - \BIGBETA * sin(\THETA / 2) * sin(\THETA / 2)))) / (\BIGBETA - 1) * \ISD}
    \pgfmathsetmacro{\BIGGAMMA}{asin(\INTERSECTION * sin(\THETA / 2) / \RADIUS)}

    \draw[draw, thick, fill=gray!50] (\INTERSECTION cm, 0)
        arc[start angle=210 + \BIGGAMMA, end angle=210 - \BIGGAMMA, radius=\RADIUS cm]
        arc[start angle=270 + \BIGGAMMA, end angle=270 - \BIGGAMMA, radius=\RADIUS cm]
        arc[start angle=-30 + \BIGGAMMA, end angle=-30 - \BIGGAMMA, radius=\RADIUS cm]
        arc[start angle=30 + \BIGGAMMA, end angle=30 - \BIGGAMMA, radius=\RADIUS cm]
        arc[start angle=90 + \BIGGAMMA, end angle=90 - \BIGGAMMA, radius=\RADIUS cm]
        arc[start angle=150 + \BIGGAMMA, end angle=150 - \BIGGAMMA, radius=\RADIUS cm];

    \draw[draw, thick, dashed] (vertex-1) -- (vertex-2) -- (vertex-3) -- (vertex-4) -- (vertex-5) -- (vertex-6) -- (vertex-1);

    \draw[draw, dotted, thick] (site-0) -- (site-1);
    \draw[draw, dotted, thick] (site-0) -- (site-6);
	\node at (0.5cm, 0) {$\theta$};
	\draw (0.25cm * \SqrtThree / 2, 0.25cm / 2) arc[start angle=\THETA / 2, end angle=-\THETA / 2, radius=0.25cm, thick];
\end{tikzpicture}
        \label{fig:multi-cells:gt}
    }
    \caption{
    The serving region of the concerned gNB with different ratios between its offset transmission power and that of its neighbors.
    $\theta$ is the angular interval between two neighbors.
    }
    \label{fig:multi-cells}
\end{figure}

Assume that a gNB has multiple neighbors distributed at the same angular interval $\theta \in (0, \pi)$, \textit{e.g.}, $\theta = 60^{\circ{}}$.
\figref{fig:multi-cells} shows its serving region under different $\beta$.
Each segment of the concerned gNB's boundary in \figref{fig:multi-cells:lt} and \ref{fig:multi-cells:gt} is a circular arc from the corresponding Apollonius circle.

\subsection{Parameters Change as a Multiplier}

When $\beta < 1$, the concerned gNB's serving region will shrink to several shaded areas in \figref{fig:bi-antenna}, where $\theta$ is the angular interval between two neighbors.
Let $\gamma$ be the corresponding central angle of the shaded area's circular arc.
$\gamma$ satisfies \eqnref{eqn:multi-antenna-boundary:gamma}.
Equation \eqnref{eqn:multi-antenna-boundary:area} further calculates the shaded area based on $\gamma$.

\begin{equation}\label{eqn:multi-antenna-boundary:gamma}
    \tan \frac{\theta}{2} = \frac{\frac{\sqrt{\beta} R}{1-\beta} \sin \gamma}{\frac{\sqrt{\beta} R}{1-\beta} \cos \gamma - \frac{\beta R}{1 - \beta}} \Rightarrow
    \gamma = \frac{\theta}{2} - \arcsin \left(\sqrt{\beta} \sin \frac{\theta}{2}\right)
\end{equation}

\begin{equation}\label{eqn:multi-antenna-boundary:area}
    \begin{aligned}
    S_{shaded} & = \gamma \left(\frac{\sqrt{\beta} R}{1 - \beta}\right)^{2} - \frac{\sqrt{\beta} R}{1-\beta} \sin \gamma \cdot \frac{\beta R}{1 - \beta} \\
	& = \frac{\left(\gamma - \sqrt{\beta} \sin \gamma\right) \beta R^{2}}{(1 - \beta)^{2}} 
    \end{aligned}
\end{equation}

We can obtain $\theta$ for \eqnref{eqn:multi-antenna-boundary:gamma} according to the deployment of gNBs.
The parameter adjustment naturally provides $\beta$ for \eqnref{eqn:multi-antenna-boundary:gamma} and \eqref{eqn:multi-antenna-boundary:area}.
However, $R$ in \eqnref{eqn:multi-antenna-boundary:area} is still unavailable for a single-cell model.
Hence, we calculate the ratio of areas after and before the adjustment, \textit{i.e.}, $\alpha(\beta \mid \theta) = S_{shaded} / \left(\frac{1}{4} R^{2} \tan \frac{\theta}{2}\right)$.
As for the situation where $\beta > 1$, we calculate $\alpha(\beta \mid \theta) = 2 - \alpha\left(\frac{1}{\beta} \mid \theta\right)$ for simplicity.

Equation \eqref{eqn:multi-antenna-boundary:area-ratio} shows the whole formula for $\alpha(\beta \mid \theta)$.
We take $\alpha(\beta \mid \theta)$ as a proxy for $\frac{S^{\prime}}{S}$, applying $\alpha(\beta \mid \theta)$ to an adjustment-free prediction for the workload, \textit{i.e.}, $\mathbf{W} \approx \tilde{\mathbf{W}} \alpha(\beta \mid \theta)$.

\begin{equation}\label{eqn:multi-antenna-boundary:area-ratio}
    \alpha(\beta \mid \theta) = \begin{cases}
        \frac{4 \beta \left(\gamma - \sqrt{\beta} \sin \gamma\right)}{(1 - \beta)^{2}\tan \frac{\theta}{2}} &, \beta < 1 \\
        1 & , \beta = 1 \\
        2 - \alpha\left(\frac{1}{\beta} \mid \theta\right) & , \beta > 1 \\
    \end{cases}
\end{equation}

\subsection{Design Overview}

\begin{figure*}[t]
    \centering

\begin{tikzpicture}[
	node distance=3cm,
	align=center,
	execute at begin node=\setlength{\baselineskip}{10pt}
	]
	\node(timestamps){$t-I, \cdots, t-1, t, \cdots, t + O - 1 $};
	\node[above of=timestamps, yshift=-2.3cm, xshift=0.1cm](parameters){$\Delta \mathbf{p}$};
	\draw[thick, ->] (parameters) -- +(0, -0.5cm);
	\node[above of=timestamps, yshift=-2cm, xshift=-2cm](data){\textit{Data}};
	\draw [thick,decorate, decoration = {brace, mirror}] (timestamps.187) -- +(2.0cm, 0);
	\draw [thick,decorate, decoration = {brace}] (timestamps.353) -- +(-2.1cm, 0);
	\node[below of=timestamps, xshift=-1.2cm, yshift=2.3cm](time-src){$\mathbf{T}_{src}$};
	\node[below of=timestamps, xshift=1.2cm, yshift=2.3cm](time-tgt){$\mathbf{T}_{tgt}$};
	\node[below of=time-src, yshift=2cm](workload-src){$\mathbf{W}_{src}$};
	\node[draw, dotted, minimum width=0.8cm, minimum height=0.5cm, below of=time-tgt, yshift=2cm](workload-tgt){};
	\node[below of=workload-src, yshift=2cm](qos-src){$\mathbf{Q}_{src}$};
	\node[draw, dotted, minimum width=0.8cm, minimum height=0.5cm, below of=workload-tgt, yshift=2cm](qos-tgt){};
	\node[below of=qos-src, yshift=2cm](other-src){$\cdots$};
	\node[below of=qos-tgt, yshift=2cm](other-tgt){$\cdots$};
	\node[draw,dashed,fit=(data) (timestamps) (other-src) (other-tgt)] {};

	\node[right of=data, xshift=3cm](architecture){\textit{\method{}}};

	\node[below of=architecture, xshift=1.5cm, yshift=2.5cm](w-historical-pa){$\mathbf{T}_{src}$};
	\node[below of=w-historical-pa, yshift=2.5cm](w-future-pa){$\mathbf{T}_{tgt}$};
	\node[below of=w-future-pa, yshift=2.5cm](w-historical-self){$\mathbf{W}_{src}$};
	\node[draw,thick, minimum width=2.5cm, minimum height=1cm, right of=w-future-pa](w-transformer){$\mathbf{W}$'s Transformer};
	\node[right of=w-transformer](w-prediction-raw){$\mathbf{W}^{\prime}_{tgt}$};
	\draw[thick, ->] (w-historical-pa) -- ++(1.3cm, 0) |-(w-transformer);
	\draw[thick, ->] (w-future-pa) -- (w-transformer);
	\draw[thick, ->] (w-historical-self) -- ++(1.3cm, 0) |-(w-transformer);
	\draw[thick, ->] (w-transformer) -- (w-prediction-raw);
	\node[above of=w-prediction-raw, yshift=-2.5cm](parameter){$\Delta \mathbf{p}$};
	\node[draw,dashed,double, right of=w-prediction-raw, xshift=-1.2cm](hsc){HSC-PEP};
	\draw[thick, ->] (w-prediction-raw) -- (hsc);
	\draw[thick, ->] (parameter) -- (hsc);
	\node[below of=hsc, yshift=2cm](w-prediction){$\tilde{\mathbf{W}}_{tgt}$};
	\draw[dashed,double, ->] (hsc) -- (w-prediction);

	\node[below of=w-historical-pa, yshift=1cm, xshift=-0.7cm](q-historical-pa){$\mathbf{W}_{src}, \mathbf{T}_{src}, \cdots$};
	\node[below of=q-historical-pa, yshift=2.5cm](q-future-pa){$\tilde{\mathbf{W}}_{tgt}, \mathbf{T}_{tgt}, \cdots$};
	\node[below of=q-future-pa, yshift=2.5cm, xshift=0.7cm](q-historical-self){$\mathbf{Q}_{src}$};
	\node[draw,thick, minimum width=2.5cm, minimum height=1cm, below of=w-transformer, yshift=1cm](q-transformer){$\mathbf{Q}$'s Transformer};
	\node[right of=q-transformer](q-prediction){$\tilde{\mathbf{Q}}_{tgt}$};
	\draw[thick, ->] (q-historical-pa) -- ++(2cm, 0) |-(q-transformer);
	\draw[thick, ->] (q-future-pa) -- (q-transformer);
	\draw[thick, ->] (q-historical-self) -- ++(1.3cm, 0) |-(q-transformer);
	\draw[thick, ->] (q-transformer) -- (q-prediction);
	\draw[dashed,double, ->] (w-prediction) -- ++(-10cm, 0) |- (q-future-pa);

	\node[below of=q-transformer, yshift=2cm]{$\cdots$};
	\node[below of=q-transformer, yshift=1.5cm](gt){\textit{Graphical Transformer}};
	\node[draw,double,dashed,fit=(w-transformer) (gt)](gt-border) {};
	\node[draw,dashed,fit=(architecture) (gt-border) (hsc)] {};
\end{tikzpicture}
    \caption{
        Architecture of \method{}.
        \method{} aims to fill the empty dotted boxes in the left part with its outputs based on the available data.
        The Graphical Transformer module is responsible for time series forecasting.
        The HSC-PEP (Homogeneous Single-Cell Parameter Effect Predictor) module converts the predicted adjustment-free workload into the one with adjustment, which is further sent to the other Transformers.
    }\label{fig:architecture}
\end{figure*}
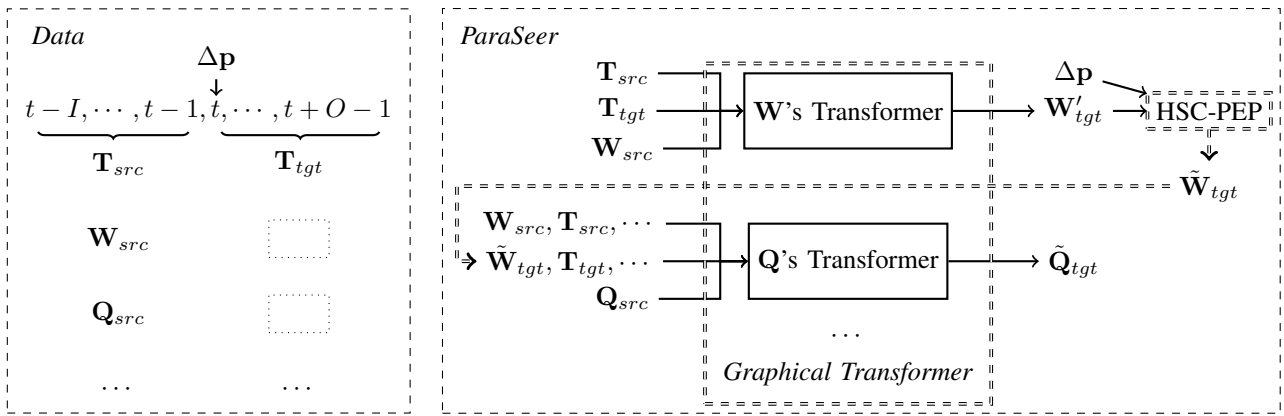

Based on our homogeneous single-cell model, we propose \textit{\method{}} to predict the parameter change's effect.
\figref{fig:architecture} presents the architecture of \method{}.
\method{} consists of several Transformers for each cluster of monitoring metrics, organizing them based on the graphical model in \figref{fig:graphical-model}, namely \textit{Graphical Transformer} (GT).
GT provides the fundamental ability of time series forecasting for \method{}.
On the other hand, we wrap $\alpha(\beta \mid \theta)$ as \textit{Homogeneous Single-Cell Parameter Effect Predictor} (HSC-PEP), adjusting GT's prediction for the workload.
GT will further propagate the parameter change's effect to other variables along the graphical model.

\section{\method{}}\label{sec:methodology}

\subsection{Graphical Transformer for Time Series Forecasting}\label{sec:methodology:gt}

\begin{figure*}[t]
    \centering

\begin{tikzpicture}[
	node distance=3cm,
	text width=2.2cm,
	align=center,
	execute at begin node=\setlength{\baselineskip}{10pt}
]
    \node[draw, minimum width=2.5cm, minimum height=0.8cm](encoder-embed) {Data Embedding};
    \node[draw, minimum width=2.5cm, minimum height=0.8cm, right of=encoder-embed](encoder) {Transformer Encoder};
    \node[draw, minimum width=2.5cm, minimum height=0.8cm, below of=encoder-embed, yshift=1cm](decoder-embed) {Data Embedding};
    \node[draw, minimum width=2.5cm, minimum height=0.8cm, right of=decoder-embed](decoder) {Transformer Decoder};
    \node[draw, minimum width=2.5cm, minimum height=0.8cm, right of=decoder](projection) {Projection};

    \node[text width=3cm, right of=projection, xshift=0.4cm](prediction) {$\hat{\mathbf{x}}^{(t)}, \cdots, \hat{\mathbf{x}}^{(t + O - 1)}$};
    \node[text width=4.4cm, left of=encoder-embed, xshift=-1.2cm, yshift=0.5cm](encoder-data) {$\mathbf{x}^{(t - I)}, \cdots, \mathbf{x}^{(t - 1)}$};
    \node[text width=4.4cm, left of=encoder-embed, xshift=-1.2cm, yshift=-0.5cm](encoder-cond) {$\mathbf{pa}^{(t - I)}(\mathbf{X}), \cdots, \mathbf{pa}^{(t - 1)}(\mathbf{X})$};
    \node[text width=4.4cm, left of=decoder-embed, xshift=-1.2cm, yshift=0.5cm](decoder-data) {$\mathbf{x}^{(t - \frac{I}{2})}, \cdots, \mathbf{x}^{(t - 1)},$ $\mathrm{average}\left\{\mathbf{x}^{(t - I)}, \cdots, \mathbf{x}^{(t - 1)}\right\}$};
    \node[text width=4.4cm, left of=decoder-embed, xshift=-1.2cm, yshift=-0.5cm](decoder-cond) {$\mathbf{pa}^{(t - \frac{I}{2})}(\mathbf{X}), \cdots, \mathbf{pa}^{(t - 1)}(\mathbf{X}),$ $\mathbf{pa}^{(t)}(\mathbf{X}), \cdots, \mathbf{pa}^{(t + O - 1)}(\mathbf{X})$};

	\draw[->, thick] (encoder-embed) -- (encoder);
	\draw[->, thick] (decoder-embed) -- (decoder);
	\draw[->, thick] (encoder) -- node[anchor=west,text width=3cm]{Cross-Attention} (decoder);
	\draw[->, thick] (decoder) -- (projection);

	\draw[->, thick] (projection) -- (prediction);
	\draw[->, thick] (encoder-data) -- +(2.5cm,0) |- (encoder-embed);
	\draw[->, thick] (encoder-cond) -- +(2.5cm,0) |- (encoder-embed);
	\draw[->, thick] (decoder-data) -- +(2.5cm,0) |- (decoder-embed);
	\draw[->, thick] (decoder-cond) -- +(2.5cm,0) |- (decoder-embed);
\end{tikzpicture}
    \caption{Transformer-based time series forecasting}
    \label{fig:forecasting:transformer}
\end{figure*}
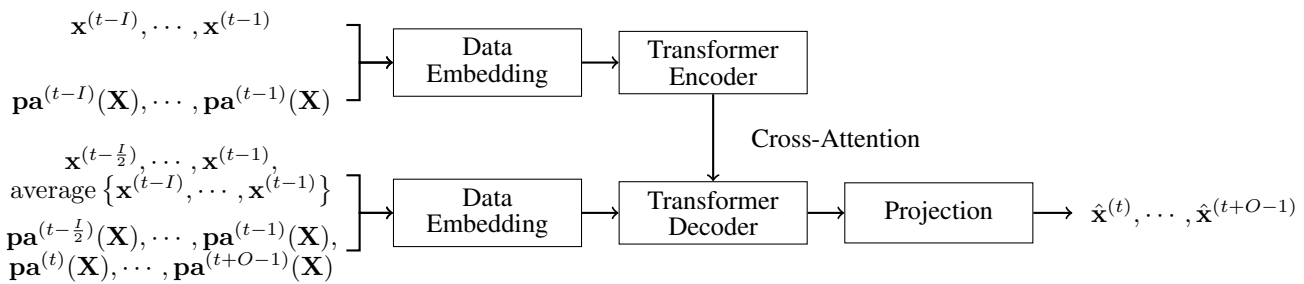

Following the existing works~\cite{Wu:2021,Zhou:2022,Nie:2023}, we adopt Transformer for time series forecasting.
For the sake of clarity, \figref{fig:forecasting:transformer} presents its architecture.
The encoder module processes the historical data and produces its hidden representation.
On the other hand, the decoder module concatenates data from two periods: 1) recent historical data for reference, and 2) an initial guess for the data of interest (\textit{e.g.}, the average of historical data) or conditions taken for ground (\textit{e.g.}, timestamps in the future).
The decoder module takes the output of the encoder one for cross-attention, calculating another hidden representation.
With a simple projection module, the final hidden representation is restored to time series of monitoring metrics, filling the initial guess with a new prediction.

Different from existing works, we organize more than one Transformer based on the graphical model, represented by $\mathbf{pa}(\mathbf{X})$ in \figref{fig:forecasting:transformer}.
The purpose of introducing the graphical model is in two folds.
On the one hand, we argue that a parameter adjustment affects variables other than the workload indirectly.
After modifying the workload by HSC-PEP, \method{} relies on the graphical model to propagate the parameter change's effect to other variables.
On the other hand, the explicit graphical model can also prevent attention on spuriously correlated data and help GT learn efficiently.
For example, interference prediction is independent of workload as encoded in the graphical model.

\subsection{Homogeneous Single-Cell Parameter Effect Predictor}

We wrap $\alpha(\beta \mid \theta)$ as an HSC-PEP module to model the effect of the adjusted transmission power and CIO, as discussed in \secref{sec:hsc}.
HSC-PEP is positioned after $\mathbf{W}$'s Transformer to adjust GT's prediction for workload, as shown in \figref{fig:architecture}.
As $\alpha(\beta \mid \theta)$ is used as a multiplier, we preprocess data with scaling only to plug HSC-PEP seamlessly into GT.
For example, the PRB uplink utilization after preprocessing will fall in $[0, 1]$ with a nonzero mean.
On the other hand, the proposed GT will take the predicted workload $\tilde{\mathbf{W}}_{tgt}$ and historical data to predict the other variables, propagating a parameter change's effect.

\subsection{Pre-training}

We pre-train each Transformer of GT with adjustment-free data on the time series forecasting task.
Mean squared error (MSE) is taken as the loss function.
Notice that data are sometimes missing.
Hence, we fill the missing points with zeros and exclude their contribution from loss~\cite{Xu:2018}, as shown in \eqnref{eqn:masked-loss}.
$N$, $S$, and $d$ represent the number of samples, the length of the time series, and the number of monitoring metrics, respectively.
$\omega_{i,j,k}$ is 0 when $x_{i,j,k}$ is missing and 1 otherwise.

\begin{equation}\label{eqn:masked-loss}
    loss = \frac{
        \sum_{i=1}^{N} \sum_{j=1}^{S} \sum_{k=1}^{d} (x_{i,j,k} - \tilde{x}_{i,j,k})^{2} \omega_{i,j,k}
    }{
        \sum_{i=1}^{N} \sum_{j=1}^{S} \sum_{k=1}^{d} \omega_{i,j,k}
    }
\end{equation}

Multi-task learning has shown its benefit in improving a model in terms of generalization~\cite{Navon:2022}.
Following existing works~\cite{Devlin:2019,He:2022}, we also consider the reconstruction task.
Specifically, the projection module in \figref{fig:forecasting:transformer} performs as a converter from hidden representation to the time series of the monitoring metrics.
We apply such a module on the encoder output to obtain the reconstruction $\tilde{\mathbf{x}}^{t-I}, \cdots, \tilde{\mathbf{x}}^{t-1}$.
Inspired by the success of masked auto-encoders in language and vision~\cite{Devlin:2019,He:2022}, we randomly mask part of $\mathbf{x}^{t-I}, \cdots, \mathbf{x}^{t-1}$ as zeros and calculate \eqnref{eqn:masked-loss} for the masked ones.

As the projection module is shared between the forecasting and reconstruction tasks, the output of the encoder module in the reconstruction task is supposed to be identical to the output of the decoder module in the forecasting task for the same sequence of the concerned time series.
We calculate MSE between these two hidden representations to formulate such a consideration.
Furthermore, both reconstruction-related losses are calculated based on the latter half of the time series, as it can be difficult to reconstruct the earlier half with little context.
We sum three losses up as the final loss function for simplicity.

\subsection{Fine-tuning}

We can fine-tune \method{} on the data with parameter adjustments to compensate for the limited diversity of pre-training data.
Notice that the parameters of HSC-PEP are independent of data.
In contrast, GT will update its parameters during fine-tuning.
In the experiment (\secref{sec:experiment:ablation:tuning}), fine-tuning brings marginal benefit for \method{} equipped with HSC-PEP.
However, it will be mandatory when HSC-PEP is replaced by a new module, which introduces new parameters.


\section{Experiments}\label{sec:experiment}

\subsection{Experimental Setup}

\subsubsection{Dataset}

We use a real-world dataset from a nationwide cellular service provider to evaluate different methods, denoted as $\mathcal{D}_{1}$.
This dataset was collected from August 11th to 30th in 2022, including 1,045 times of cell parameter adjustments with distinct cells in the same geographical region.
There are, on average, 628 data points per case before an adjustment.
The sampling interval is 15 minutes.
We filter in 17 monitoring metrics falling in the 3 clusters shown in \tableref{tab:variables}, besides 3 time-related features.
We postpone the introduction of another dataset $\mathcal{D}_{2}$ in \secref{sec:experiment:new-region}, which is used to verify the generalization for different geographical regions.

\subsubsection{Evaluation Metrics}

We measure the quality of the predicted time series for each monitoring metric separately to provide a holistic evaluation.
Based on the original value before preprocessing, we adopt \textit{root mean squared error} (RMSE) and \textit{symmetric mean absolute percentage error} (sMAPE) to represent the distance between true observations and predicted time series~\cite{Jiang:2022}, as shown in \eqnref{eqn:evaluation-metric:rmse} and \eqref{eqn:evaluation-metric:smape}, respectively.
Compared with MAPE, sMAPE takes the average of true observations and predicted values as the divisor to handle zeros, \textit{e.g.}, the drop rate is zero almost all the time.

\begin{equation}\label{eqn:evaluation-metric:rmse}
    RMSE = \sqrt{\underset{i,j,k}{\mathrm{average}} \left(x_{i,j,k} - \tilde{x}_{i,j,k}\right)^{2}}
\end{equation}

\begin{equation}\label{eqn:evaluation-metric:smape}
    sMAPE = \underset{i,j,k}{\mathrm{average}} \left| \frac{x_{i,j,k} - \tilde{x}_{i,j,k}}{\left(x_{i,j,k} + \tilde{x}_{i,j,k}\right) / 2} \right|
\end{equation}

In the experiments, we randomly sample 50\% of cases for testing and the rest for training, repeated 10 times for fair comparison.
We calculate the averaged evaluation metrics over testing cases for each monitoring metric in each trial.
The rest of this section will present the average of 10 trials.

\subsubsection{Baselines}

We select a well-known machine learning model and two causal effect estimation methods in the recent literature as baseline methods.

\begin{itemize}
    \item Elsayed et al. find that Gradient Boosting Regression Tree (\textit{GBRT}) is competitive with deep learning models for time series forecasting~\cite{Elsayed:2021}.
    For each monitoring metric, a GBRT is trained with the latest values of the other monitoring metrics before the adjustment as the input, besides a historical sequence of the concerned time series.
    Furthermore, we enrich the input vector with $\Delta \mathbf{p}$.
    \item Varying Coefficient Neural Network (\textit{VCNet})~\cite{Nie:2021} treats the parameters of a Multilayer Perceptron (MLP) as different functions of the intervention (\textit{i.e.}, $\Delta \mathbf{p}$ in this work).
    As a result, the intervention will not be concatenated with features, emphasizing the impact of the former.
    To apply VCNet for our task, we take the historical sequence as the input (covariates) and time series in the future as the output (causal effect).
    \item \textit{$\beta$-Intact-VAE}~\cite{Wu:2022} follows the structure of the Variational Autoencoder (VAE)~\cite{Kingma:2014}.
    It models a prognostic score with a latent variable, alleviating the limited covariate overlap.
    The time series before the adjustment is taken as the input, which is the same as VCNet.
\end{itemize}

Moreover, we compare \method{} with several variants to show different components' effects.
To illustrate the effect of HSC-PEP, we replace HSC-PEP with \textit{Multiplier} and VCNet, denoted as \textit{\method{} w Multiplier} and \textit{\method{} w VCNet}, respectively.
Different from HSC-PEP, \textit{Multiplier} adopts a data-driven manner to learn the ratio between workload with and without an adjustment.
What's more, \textit{Multiplier} is enhanced by the feature extracted from historical sequence by Gated Recurrent Unit~\cite{Chung:2014} to address unobserved usage density.
\textit{\method{} w/o recons} removes the reconstruction task from pre-training.
\method{} skips the fine-tuning stage by default while \textit{\method{} w tuning} further tunes GT combined with HSC-PEP.

\subsubsection{Implementation}

In this work, we are interested in predicting the next day.
Hence, we choose 96 for $O$, which spans one day in our datasets.
On the other hand, we also set $I$ as 96, as one day can provide periodic characteristics~\cite{Li:2022}.

VCNet and $\beta$-Intact-VAE are trained by an AdamW optimizer~\cite{Loshchilov:2019} for 250 epochs with a batch size of 16.
The learning rate is set empirically as $10^{-3}$.
Moreover, we set the learning rate as $10^{-4}$ for warm-up in the first 10 epochs.
After 100 epochs, we discount the learning rate by 0.97 after each epoch.

As for \method{}, the pre-training stage adopts the AdamW optimizer~\cite{Loshchilov:2019} for 2 epochs with a batch size of 32 and a learning rate of $10^{-4}$.
We use the historical data before adjustments of all the 1,045 cases in pre-training.
We will discuss the performance of \method{} on new cells by pre-training on the training cases only in \secref{sec:experiment:ablation:new-cell}.
In the fine-tuning stage, \textit{\method{} w tuning} is further trained for 5 epochs with a batch size of 16 and a learning rate of $10^{-4}$.
Because \textit{\method{} w Multiplier} and \textit{\method{} w VCNet} introduce new parameters, we first freeze the parameters of GT and train the corresponding Multiplier or VCNet for 100 epochs with the learning rate of $10^{-4}$ for 10 epochs and $10^{-3}$ for the rest.
After that, these two variants are further tuned for 5 epochs in the same way as \textit{\method{} w tuning}.

\subsection{Performance Evaluation}

\begin{figure}[t]
    \centering
    \includegraphics[width=0.98\columnwidth]{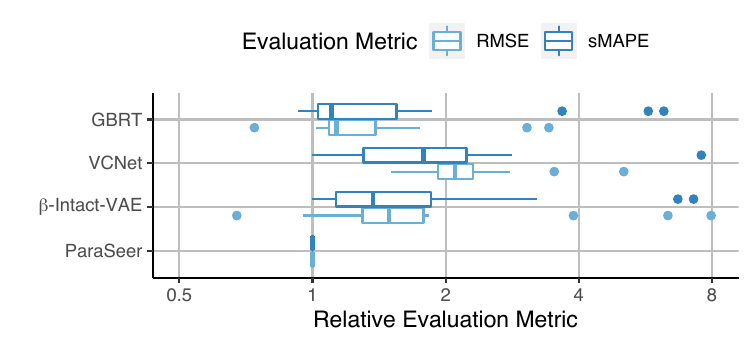}
    \caption{Performance evaluation of \method{} and baseline methods on $\mathcal{D}_{1}$. The axes for evaluation metrics are logarithmically scaled.}\label{fig:experiment:comparison}
\end{figure}

\figref{fig:experiment:comparison} compares \method{} with GBRT~\cite{Elsayed:2021}, VCNet~\cite{Nie:2021}, and $\beta$-Intact-VAE~\cite{Wu:2022}.
We present the evaluation results with box plots to show the performance of each method for different monitoring metrics.
Three vertical lines of each box represent the lower quartile, the median, and the upper quartile of evaluation metrics for different monitoring metrics.
For ease of understanding, we draw each box with relative evaluation metrics with those of \method{} as 1.
As a result, a box on the right of \method{}'s box implies that \method{} outperforms the corresponding baseline significantly.

As shown in \figref{fig:experiment:comparison}, \method{} outperforms baselines in terms of RMSE and sMAPE.
\tableref{tab:experiment:comparison} further presents the evaluation results averaged among monitoring metrics for clarity.
As shown in \tableref{tab:experiment:comparison}, \method{} improves RMSE and sMAPE by 25.8\% and 59.0\%, compared with the best-performed baseline.
The simple GBRT method is more effective than two deep learning-based baselines.
Such a phenomenon has been observed in some previous research~\cite{Elsayed:2021,Zeng:2023}.
Thanks to the detailed modeling provided by GT and HSC-PEP, \method{} achieves significant progress compared with GBRT.

\begin{table}[t]
    \centering
	\begin{threeparttable}[c]
    \caption{
        Performance evaluation of \method{} and baseline methods on $\mathcal{D}_{1}$, averaged among monitoring metrics.
    }\label{tab:experiment:comparison}
    \begin{tabular}{lrcrc}
        \toprule
        \multirow{2}{*}{Method} & \multicolumn{2}{c}{Relative RMSE} & \multicolumn{2}{c}{Relative sMAPE} \\
        & mean (std) & p-value & mean (std) & p-value \\
        \midrule
        GBRT & 1.347(0.80) & 0.007\tnote{*} & 2.440(2.69) & 0.000\tnote{*} \\ 
        VCNet & 6.739(9.53) & 0.000\tnote{*} & 5.220(11.5) & 0.000\tnote{*} \\ 
        $\beta$-Intact-VAE & 4.996(10.3) & 0.002\tnote{*} & 11.693(37.5) & 0.000\tnote{*} \\ 
        \method{} & 1.000(0.00) & / & 1.000(0.00) & / \\ 
        \bottomrule
    \end{tabular}
	\begin{tablenotes}
		\item [*] statistically significant as $p < 0.05$
	\end{tablenotes}
	\end{threeparttable}
\end{table}

The performance of a method varies with monitoring metrics, as shown in \figref{fig:experiment:comparison}, indicating the diversity among monitoring metrics.
We argue that a single statistical quantity like the mean and the standard deviation is insufficient to measure a method properly.
Hence, we choose the box plot to present the results in \figref{fig:experiment:comparison}.
Moreover, we conduct Wilcoxon signed rank tests for ``the relative RMSE (sMAPE) is larger than 1'' (\textit{i.e.}, ``\method{} achieves smaller RMSE (sMAPE) than a given baseline''), which is represented by the \textit{p-value} columns in \tableref{tab:experiment:comparison}.
In summary, \method{} outperforms GBRT, VCNet, and $\beta$-Intact-VAE statistically significantly in terms of both RMSE and sMAPE.

\subsection{Ablation Studies}

To show the effect of \method{}'s components, we conduct ablation studies, comparing \method{} with its variants.
Because HSC-PEP is designed for workload, the graphical model is mandatory for \method{} to propagate a parameter change's effect from the workload to other variables.
As a result, we will not remove the graphical model from \method{}.
On the other hand, HSC-PEP works as a multiplier, relying on GT's forecasting.
Hence, pre-training cannot be skipped, either.

\begin{figure}[t]
    \centering
    \includegraphics[width=0.98\columnwidth]{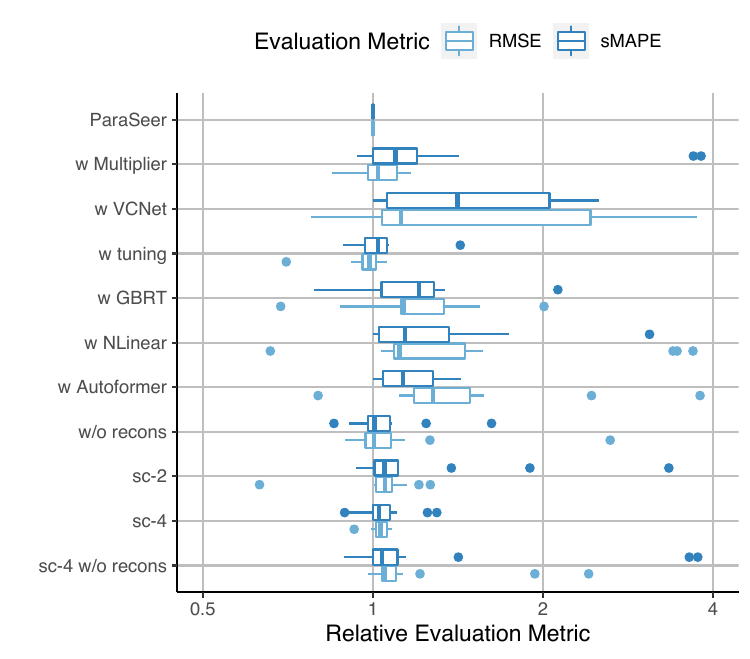}
    \caption{Performance evaluation of \method{}'s variants on $\mathcal{D}_{1}$. The axes for evaluation metrics are logarithmically scaled.}\label{fig:experiment:ablation}
\end{figure}

\begin{table}[t]
    \centering
	\begin{threeparttable}[c]
    \caption{
        Performance evaluation of \method{}'s variants on $\mathcal{D}_{1}$, averaged among monitoring metrics.
    }\label{tab:experiment:ablation}
    \begin{tabular}{crcrc}
        \toprule
        \multirow{2}{*}{Variant} & \multicolumn{2}{c}{Relative RMSE} & \multicolumn{2}{c}{Relative sMAPE} \\
        & mean (std) & p-value & mean (std) & p-value \\
        \midrule
        \method{} & 1.000(0.00) & / & 1.000(0.00) & / \\
        \midrule
        w Multiplier & 0.973(0.26) & 0.189 & 1.409(0.89) & 0.001\tnote{*} \\ 
        w VCNet & 1.621(0.99) & 0.003\tnote{*} & 2.005(1.34) & 0.000\tnote{*} \\
        w tuning & 0.919(0.23) & 0.934 & 1.474(1.27) & 0.112 \\ 
        \midrule
        w GBRT & 1.458(1.33) & 0.013\tnote{*} & 2.007(1.80) & 0.000\tnote{*} \\ 
        w NLinear & 5.219(13.7) & 0.004\tnote{*} & 13.400(44.9) & 0.000\tnote{*} \\ 
        w Autoformer & 16.179(49.8) & 0.001\tnote{*} & 36.496(133.) & 0.000\tnote{*} \\ 
        \midrule
        w/o recons & 1.082(0.43) & 0.356 & 1.957(2.02) & 0.040\tnote{*} \\ 
        sc-2 & 1.049(0.13) & 0.002\tnote{*} & 1.442(0.98) & 0.001\tnote{*} \\ 
        sc-4 & 0.982(0.21) & 0.017\tnote{*} & 1.464(1.19) & 0.017\tnote{*} \\ 
        sc-4 w/o recons & 1.193(0.38) & 0.000\tnote{*} & 1.637(1.39) & 0.013\tnote{*} \\ 
        \bottomrule
    \end{tabular}
	\begin{tablenotes}
		\item [*] statistically significant as $p < 0.05$
	\end{tablenotes}
	\end{threeparttable}
\end{table}

\subsubsection{Effect of HSC-PEP}

As shown in \figref{fig:experiment:ablation}, \method{} outperforms two variants, \textit{\method{} w Multiplier} and \textit{\method{} w VCNet}, illustrating the positive effect of HSC-PEP.
Compared with VCNet, HSC-PEP brings 38.3\% and 50.1\% improvement in RMSE and sMAPE, respectively.
Multiplier follows the same core idea of HSC-PEP except that the former learns the mapping in a data-driven manner.
As shown in \tableref{tab:experiment:ablation}, HSC-PEP helps \method{} achieve a competitive RMSE and a lower sMAPE than \textit{\method{} w Multiplier}, implying the value of domain knowledge.
On the other hand, the insight behind HSC-PEP helps \textit{\method{} w Multiplier} outperform GBRT and \textit{\method{} w VCNet}.

\subsubsection{Effect of Fine-Tuning}\label{sec:experiment:ablation:tuning}

As shown in \figref{fig:experiment:ablation}, the performance of \method{} (without tuning) is competitive with its variant with tuning.
The Wilcoxon signed rank test for ``\textit{\method{} w tuning} achieves smaller RMSE (sMAPE) than \method{} for different monitoring metrics'' provides a p-value of 0.073 (0.897), which is not statistically significant, either.
The results imply that the derived formula for $\alpha(\beta \mid \theta)$ is representative enough for the data with parameter adjustment, helping \method{} eliminate the thirst for data.

\subsubsection{Effect of Transformer}

The Transformer is prevalent in the recent time series forecasting literature, while its performance is still under doubt~\cite{Zeng:2023}.
To illustrate Transformer's effect on \method{}, we replace a Transformer with GBRT~\cite{Elsayed:2021}, NLinear~\cite{Zeng:2023} (one linear layer of neural network), and Autoformer~\cite{Wu:2021}, respectively.
The performance degradations of \textit{\method{} w GBRT} and \textit{\method{} w NLinear}, as shown in \figref{fig:experiment:ablation} and \tableref{tab:experiment:ablation}, empirically imply the superiority of Transformer-based time series forecasting.
Though Autoformer~\cite{Wu:2021} enhances Transformer with time series decomposition, \textit{\method{} w Autoformer} also underperforms \method{}.
This phenomenon can be attributed to the poor smoothness and periodicity of the cellular traffic, as shown in \figref{fig:example}.
Moreover, decomposition may lead to a biased trend for time series that is zero almost all the time, \textit{e.g.}, drop rates and congestion rates in $\mathcal{D}_{1}$.

\subsubsection{Generalization for Unseen Cells}\label{sec:experiment:ablation:new-cell}

Unseen cells' data pattern novelty threats the generalization of a data-driven method.
To explore \method{}'s performance for unseen cells, we introduce two more variants, pre-training only on the same cases as the training ones.
In other words, these variants will not access the historical data from the testing cases.
Meanwhile, a reduction in data reduces the number of batches with the same training epochs.
Hence, we pre-train these two variants with 2 and 4 epochs, respectively.
As a result, the variant pre-training 4 epochs shares a similar batch number as \method{}.
Denote the two variants as \textit{\method{} sc-n}, where \textit{sc} represents the \textbf{s}ame \textbf{c}ases and \textit{n} is the number of pre-training epochs, \textit{i.e.}, 2 and 4.

Intuitively, \textit{\method{} sc-2} and \textit{\method{} sc-4} underperform \method{} due to the data reduction, as shown in \figref{fig:experiment:ablation}.
Hence, we suggest bringing as many historical data into the pre-training process as possible to enrich data diversity.
Fortunately, the performance degradation is marginal, as shown in \figref{fig:experiment:ablation} and \tableref{tab:experiment:ablation}.

On the other hand, we introduce the reconstruction task during pre-training because we believe an extra restriction can alleviate the risk of over-fitting.
It seems that the extra task brings marginal improvement for \method{} (\textit{\method{} w/o recons} in \figref{fig:experiment:ablation}).
As we remove the reconstruction task from \textit{\method{} sc-4}, \textit{\method{} sc-4 w/o recons} performs worse than \textit{\method{} sc-4}.
We conclude that the extra reconstruction task alleviates the performance degradation due to the data reduction.
In summary, multi-task learning may be redundant for seen cells but can be beneficial for unseen ones.

\subsubsection{Hyperparameter Sensitivity}

\begin{figure}[t]
    \centering
    \includegraphics[width=0.98\columnwidth]{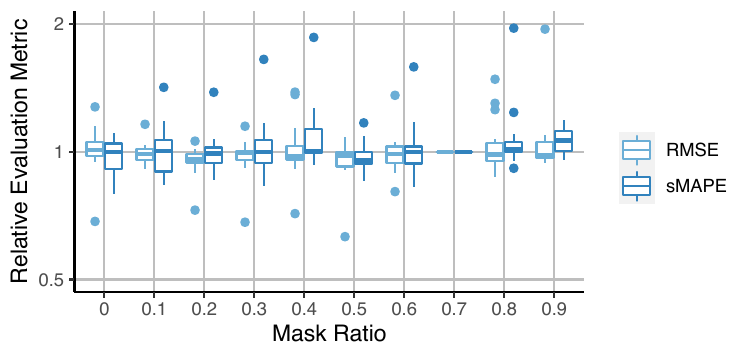}
    \caption{Performance evaluation of \method{} with different mask ratios on $\mathcal{D}_{1}$. The axes for evaluation metrics are logarithmically scaled.}\label{fig:experiment:hyperparameter}
\end{figure}

\figref{fig:experiment:hyperparameter} varies the mask ratio in the reconstruction task for \method{}.
A mask ratio of 0 means that the MSE between two kinds of hidden representations (\textit{i.e.}, for the reconstruction task and the forecasting one) is calculated while the reconstruction loss is zero.
\method{} seems not sensitive to the mask ratio.
We choose a mask ratio of 0.7 as the default, providing a relatively hard task for representation learning.

\subsection{Generalization for Cells in Another Region}\label{sec:experiment:new-region}

In this part, we further explore \method{}'s performance for cells in geographical regions different from that of $\mathcal{D}_{1}$.
We collect 63 more cases from another region, denoted as $\mathcal{D}_{2}$.
Cell parameters of each case were adjusted on May 5th, 2022.
gNBs in the two regions of $\mathcal{D}_{1}$ and $\mathcal{D}_{2}$ are operated by two different groups of operators.
We manage to map the monitoring metrics of $\mathcal{D}_{2}$ into the 17 ones of $\mathcal{D}_{1}$.
As the paging congestion rate is missing in $\mathcal{D}_{2}$, we fill this monitoring metric as zero for $\mathcal{D}_{2}$ and drop it during evaluation.

We train GBRT~\cite{Elsayed:2021}, VCNet~\cite{Nie:2021}, and $\beta$-Intact-VAE~\cite{Wu:2022} with all 1,045 cases in $\mathcal{D}_{1}$, evaluating them in $\mathcal{D}_{2}$.
Note that we already pre-train \method{} with the adjustment-free period of each case in  $\mathcal{D}_{1}$.
\method{} maintains the advantage over baseline methods on $\mathcal{D}_{2}$ statistically significantly in terms of both RMSE and sMAPE, as shown in \tableref{tab:experiment:comparison:sd}.

\begin{table}[t]
	\centering
	\begin{threeparttable}[c]
    \caption{
        Performance evaluation of \method{} and baseline methods on $\mathcal{D}_{2}$, averaged among monitoring metrics.
    }\label{tab:experiment:comparison:sd}
    \begin{tabular}{lrcrc}
        \toprule
        \multirow{2}{*}{Method} & \multicolumn{2}{c}{Relative RMSE} & \multicolumn{2}{c}{Relative sMAPE} \\
        & mean (std) & p-value & mean (std) & p-value \\
        \midrule
        GBRT & 1.665(0.76) & 0.001\tnote{*} & 3.294(7.88) & 0.002\tnote{*} \\ 
        VCNet & 7.523(14.5) & 0.000\tnote{*} & 10.313(19.7) & 0.000\tnote{*} \\ 
        $\beta$-Intact-VAE & 7.259(10.7) & 0.000\tnote{*} & 14.127(27.6) & 0.000\tnote{*} \\ 
        \method{} & \textbf{1.000}(0.00) & / & \textbf{1.000}(0.00) & / \\
        \bottomrule
    \end{tabular}
    \begin{tablenotes}
        \item [*] statistically significant as $p < 0.05$
    \end{tablenotes}
	\end{threeparttable}
\end{table}

\method{} achieves, on average, an RMSE of 6.07 for PRB downlink utilization (\%) on $\mathcal{D}_{1}$, with an even lower RMSE of 5.72 on $\mathcal{D}_{2}$.
Meanwhile, \method{}'s RMSE for the maximum number of RRC connections is 7.65 on $\mathcal{D}_{1}$ but 13.59 on $\mathcal{D}_{2}$.
The results imply the differences between the two datasets, which need further identification.
Fortunately, the Wilcoxon signed rank test for ``\method{}'s relative RMSE (sMAPE) on $\mathcal{D}_{2}$ is larger than 1'' provides a p-value of 0.174 (0.812), implying \method{}'s application potential for other regions.

\begin{figure}[t]
    \centering
    \includegraphics[width=0.95\columnwidth]{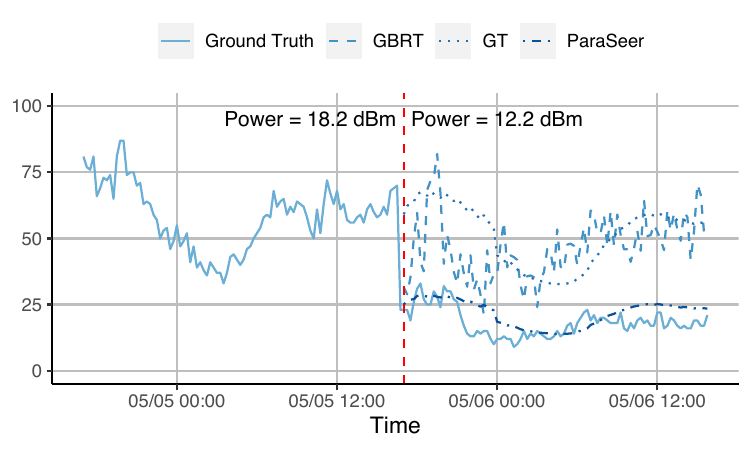}
    \caption{
        The maximum number of RRC connections in a real-world case from $\mathcal{D}_{2}$, enriching \figref{fig:example} with predictions.
    }\label{fig:experiment:case}
\end{figure}

We further present the effect of \method{} with a case from $\mathcal{D}_{2}$.
As shown in \figref{fig:experiment:case}, GT predicts a time series above the real one without noticing parameter adjustment (the dotted line).
The effect of HSC-PEP is to pull GT's prediction closer to the ground truth time series (the darkest line).
We plan to enhance GT's forecasting accuracy for better prediction.
On the other hand, GBRT predicts a time series (the dashed line)  close to GT's prediction, underestimating the parameter change's effect.

\section{Related Works}\label{sec:relate-works}

\noindent
\textbf{Cellular Traffic Prediction.}
ABSENCE~\cite{Nguyen:2015} and TOIP~\cite{Yang:2017} count the historical traffic for each time window (\textit{e.g.}, the same day of the week) for each region.
Thus, the traffic in the future is modeled by the learned distribution.
Such coarse-grained modeling neglects the correlation of a consecutive time series.
In contrast, dynamical traffic prediction for the near future enables proactive network optimization, \textit{e.g.}, load balancing~\cite{Wang:2022}.
Graph neural networks~\cite{Wang:2017,Wang:2021,Zhou:2022:traffic} are popular in the recent literature, capturing the spatial-temporal correlation among traffic.
We refer readers to a recent survey~\cite{Jiang:2022} for a thorough discussion.

\noindent
\textbf{Time Series Forecasting.}
Traditional methods (\textit{e.g.}, ARIMA~\cite{Wang:2022:DARIMA}) model an arbitrary data generation process with a linear combination of several historical data points in the near past.
Prophet~\cite{Taylor:2018} decomposes time series into the trend, seasonality, and holidays, forecasting each part separately.
With the success of Transformer~\cite{Vaswani:2017} in language and vision, some recent works develop Transformer-based deep learning methods for time series forecasting~\cite{Wu:2021,Zhou:2022,Nie:2023,Zhou:2021}.
The works mentioned above, including the cellular traffic prediction literature, imply an implicit assumption that the data generation mechanism is stable.
In other words, there is no intervention~\cite{Bareinboim:2022} to the system (\textit{i.e.}, parameter adjustments discussed in this work).

\noindent
\textbf{Causal Inference.}
The fundamental challenge of causal effect estimation from observational data~\cite{Zhang:2021:TESurvey} is that the same unit cannot be intervened in two different ways.
The causal forest~\cite{Wager:2018} groups the training samples with a set of splitting criteria.
Samples in a leaf are taken as data from a random controlled trial, assuming that the leaf is small enough.
Such an assumption requires a large quantity of data, failing to predict unseen data points.
Deep learning-based methods extend the single-model approaches~\cite{Zhang:2021:TESurvey}, emphasizing the effect of interventions~\cite{Nie:2021} or explicitly modeling unobserved variables~\cite{Zhang:2021,Wu:2022}.
Unfortunately, it is hard for a deep learning-based method to learn the same latent distribution as the underlying data generation process~\cite{Wu:2022}.
Hence, \method{} integrates domain knowledge explicitly as HSC-PEP to alleviate the model error.
Readers can find more information on causal effect estimation in a recent survey~\cite{Zhang:2021:TESurvey}.


\section{Conclusion}\label{sec:conclusion}

Configuration and optimization parameters, like the transmission power and CIO, provide flexible options for operators to optimize the cellular network.
Predicting a parameter change's effect can prevent an improper parameter adjustment.
In this work, we focus on predicting the status (like the workload and QoS) of a single cell after a parameter adjustment.
The analysis will be conducted before an adjustment is applied to provide an early inspection.

To address this problem, we propose \method{}, decomposing the prediction into an adjustment-free time series forecasting task and modifying the adjustment-free prediction.
We develop a homogeneous single-cell model, based on which a formula is derived for calculating the ratio of cell areas after and before an adjustment from the underlying mechanism of the transmission power and CIO.
\method{} applies the calculated ratio as a multiplier to the predicted workload, further propagating the parameter change's effect to other variables via a graphical model.
On the other hand, \method{} utilizes plenty of adjustment-free data to pre-train a Graphical Transformer for the time series forecasting task.
\method{} outperforms the best baseline by more than 25.8\% in RMSE on two real-world datasets.
Ablation studies further illustrate the contribution of \method{}'s components and the generalization for unseen cells.

In the future, we plan to extend \method{} for multiple cells as a single-cell model knows insufficient information.
Thus, efficiency for both data processing and inference may become an obstacle for a large cellular network.
Furthermore, the follow-up work with a precise multi-cell model is to recommend parameters for the desired status.

\section*{Acknowledgment}

We thank Zhe Xie and Pengtian Zhu for their helpful discussions.
The experiments are supported by JIUTIAN Artificial Intelligence Platform.
This work is supported by the National Key Research and Development Program of China (No.2019YFE0105500), the Research Council of Norway (No.309494), the State Key Program of National Natural Science of China under Grant 62072264, and the Beijing National Research Center for Information Science and Technology (BNRist) key projects.


\appendix[Discussion on \thmref{thm:track-necessity}]

\label{sec:appendix:track-necessity}

\begin{proof}[Proof of \thmref{thm:track-necessity}]

Notice that $\lim_{\phi_{i} \to 0^{+}} \mathbf{S}_{i} = \{\mathbf{c}_{i}\}$.
Hence, we can calculate $\rho$ for each site point with $\rho(c_{i,1}, c_{i, 2}) = \lim_{\phi_{i} \to 0^{+}} \frac{f_{i}(\Phi)}{\lVert \mathbf{S}_{i} \rVert}$, where $\lVert \mathbf{S}_{i} \rVert$ is the area of the region $\mathbf{S}_{i}$ and $\lim_{\phi_{i} \to 0^{+}} \lVert \mathbf{S}_{i} \rVert = 0$.

The following analysis starts with equal weight, \textit{i.e.}, $\phi_{i} = \phi_{0}, i = 1, 2, \cdots$.
We will reconstruct the density $\rho(x_{1}, x_{2})$ for an arbitrary region $\mathbf{S}_{i}$ and a point other than the site point ($\mathbf{x} \in \mathbf{S}_{i}, \mathbf{x} \neq \mathbf{c}_{i}$) with three steps.

\paragraph{Shrink $\mathbf{S}_{i}$, making $\mathbf{x}$ at the boundary}

When we decrease $\phi_{i}$ from $\phi_{0}$ to $0$, $\mathbf{S}_{i}$ will shrink to $\{\mathbf{c}_{i}\}$, no longer including $\mathbf{x}$.
During the continuous change of $\phi_{i}$, there exists a value $\phi_{i}^{\prime} \in (0, \phi_{0})$ that makes $\mathbf{x}$ located at the boundary of the new region generated by $\mathbf{c}_{i}$.

\paragraph{Enlarge a neighboring region, making $\mathbf{x}$ at the vertex}

After we change $\phi_{i}$ from $\phi_{0}$ to $\phi_{i}^{\prime}$, $\mathbf{x}$ is located at the border of two regions.
Denote the other region as $\mathbf{S}_{j}, j \neq i$.
Let $\mathbf{S}_{k}, k \notin \{i, j\}$ be a neighbor of $\mathbf{S}_{i}$, intersecting with $\mathbf{S}_{j}$ at $\mathbf{S}_{i}$'s vertex.
When we increase $\phi_{k}$ from $\phi_{0}$ to $+\infty$, $\mathbf{S}_{k}$ will be enlarged and finally include $\mathbf{x}$.
Hence, there exists a value $\phi_{k}^{\prime} \in [\phi_{0}, +\infty)$ that makes $\mathbf{x}$ located at the border between $\mathbf{S}_{i}$ and $\mathbf{S}_{k}$, where $\frac{\lVert \mathbf{x} - \mathbf{c}_{k} \rVert}{\phi_{k}^{\prime}} = \frac{\lVert \mathbf{x} - \mathbf{c}_{i} \rVert}{\phi_{i}^{\prime}} = \frac{\lVert \mathbf{x} - \mathbf{c}_{j} \rVert}{\phi_{0}}$.
In other words, $\mathbf{x}$ is now located at the vertex of $\mathbf{S}_{i}$.

Notice that the non-existence of $\mathbf{S}_{k}$ will imply the border between $\mathbf{S}_{i}$ and $\mathbf{S}_{j}$ to be a straight line.
As a result, $\lVert \mathbf{S}_{i} \rVert$ will be infinite, which violates the $\lVert \mathbf{S}_{i} \rVert < +\infty$ restriction.

\paragraph{Calculate the difference in difference}

Increasing $\phi_{j} = \phi_{0}$ (or $\phi_{k}^{\prime} \ge \phi_{0}$) by a small enough positive value $\Delta \phi_{j}$ (or $\Delta \phi_{k}$) will make $\mathbf{S}_{i}$'s corresponding border shrink to a dashed arc in \figref{fig:appendix:difference-in-difference}.
The shaded region in \figref{fig:appendix:difference-in-difference} includes the concerned $\mathbf{x}$.
Equation~\eqref{eqn:appendix:difference-in-difference} calculates the quantity $\Delta$ in the shaded region.

\begin{equation}\label{eqn:appendix:difference-in-difference}
    \begin{aligned}
    \Delta = & \overbrace{\left[
        f_{i}(\phi_{j}, \phi_{k}^{\prime}, \cdots) - f_{i}(\phi_{j} + \Delta \phi_{j}, \phi_{k}^{\prime}, \cdots)
    \right]}^{\text{difference brought by the change in $\phi_{j}$}} - \\
    & \underbrace{\left[
    f_{i}(\phi_{j}, \phi_{k}^{\prime} + \Delta \phi_{k}, \cdots) - f_{i}(\phi_{j} + \Delta \phi_{j}, \phi_{k}^{\prime} + \Delta \phi_{k}, \cdots)
    \right]}_{\text{quantity for the upper right strip of the shaded region}}
    \end{aligned}
\end{equation}

\begin{figure}[h]
    \centering

\begin{tikzpicture}[
	align=center,
	execute at begin node=\setlength{\baselineskip}{10pt}
	]
    \input{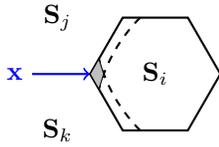}

    \node at (site-0) {$\mathbf{S}_{i}$};
    \node at (site-3) {$\mathbf{S}_{j}$};
    \node at (site-4) {$\mathbf{S}_{k}$};
    \draw[thick, <-, blue] (vertex-4) -- +(-1cm, 0) node[fill=white] {$\mathbf{x}$};

    \pgfmathsetmacro{\BIGBETA}{1 / \BETA}
    \pgfmathsetmacro{\INTERSECTION}{(\BIGBETA * cos(\THETA / 2) - sqrt(\BIGBETA * (1 - \BIGBETA * sin(\THETA / 2) * sin(\THETA / 2)))) / (\BIGBETA - 1) * \ISD}
    \pgfmathsetmacro{\BIGGAMMA}{asin(\INTERSECTION * sin(\THETA / 2) / \RADIUS)}

    \draw[thick, dashed, name path=arc-j] (-\BEL cm * 1.5 + \INTERSECTION cm, \BEL cm * \SqrtThree / 2)
        arc[start angle=150 - \BIGGAMMA, end angle=150 + \BIGGAMMA, radius=\RADIUS cm]
        coordinate (arc-j-end);
    \draw[thick, dashed, name path=arc-k] (-\BEL cm * 1.5 + \INTERSECTION cm, -\BEL cm * \SqrtThree / 2)
        arc[start angle=210 + \BIGGAMMA, end angle=210 - \BIGGAMMA, radius=\RADIUS cm]
        coordinate (arc-k-end);
    \path [name intersections={of=arc-j and arc-k,by=arc-intersection}];

    \draw[draw, thick] (vertex-1) -- (vertex-2) -- (vertex-3) -- (vertex-4) -- (vertex-5) -- (vertex-6) -- (vertex-1);

    \draw[fill=gray!50] (vertex-4) -- (arc-j-end) -- (arc-intersection) -- (arc-k-end) -- (vertex-4);
\end{tikzpicture}
    \caption{
        Isolate the neighborhood (the shaded region) of $\mathbf{S}_{i}$'s vertex $\mathbf{x}$ by manipulating the weights of $\mathbf{S}_{j}$ and $\mathbf{S}_{k}$.
    }
    \label{fig:appendix:difference-in-difference}
\end{figure}

Denote the area of the shaded region as $\Delta_{S}$.
We can reconstruct the density with $\rho(x_{1}, x_{2}) = \lim_{\Delta \phi_{j}, \Delta \phi_{k} \to 0^{+}} \frac{\Delta}{\Delta_{S}}$.

\end{proof}

With the density function reconstructed from $F$, we can filter in points with positive density to form the user trajectories.
Notice that \thmref{thm:track-necessity} aims to point out the conflict between perfect prediction and user privacy.
Practical reconstruction steps fall out of the scope of this work.

\bibliographystyle{IEEEtran}
\bibliography{IEEEabrv,bibliography}

\end{document}